\documentclass[
    ,final            
  ]
  {aipproc}

\layoutstyle{6x9}

\usepackage{dsfont}
\usepackage{psfrag}
\usepackage{amsmath,amssymb}

\newtheorem{theorem}{Theorem}
\newtheorem{sloppytheorem}{Sloppy theorem}

\begin{document}

\title{Stationary points approach to\\ thermodynamic phase transitions}


\author{Michael Kastner}{
  address={National Institute for Theoretical Physics (NITheP), Stellenbosch 7600, South Africa, and\\ Institute of Theoretical Physics,  University of Stellenbosch, Stellenbosch 7600, South Africa}
}

\begin{abstract}
Nonanalyticities of thermodynamic functions are studied by adopting an approach based on stationary points of the potential energy. For finite systems, each stationary point is found to cause a nonanalyticity in the microcanonical entropy, and the functional form of this nonanalytic term is derived explicitly. With increasing system size, the order of the nonanalytic term grows, leading to an increasing differentiability of the entropy. It is found that only ``asymptotically flat'' stationary points may cause a nonanalyticity that survives in the thermodynamic limit, and this property is used to derive an analytic criterion establishing the existence or absence of phase transitions. We sketch how this result can be employed to analytically compute transition energies of classical spin models.
\end{abstract}

\maketitle

A phase transition is an abrupt change of the macroscopic properties of a many-particle system under variation of a control parameter. An approach commonly used for the theoretical description of phase transitions is the investigation of the analyticity properties of thermodynamic functions like the canonical free energy of enthalpy. It is long known that nonanalytic behavior in a canonical or grandcanonical thermodynamic function can occur only in the thermodynamic limit in which the number of degrees of freedom $N$ of the system goes to infinity \cite{Griffiths}. Recently, however, it was observed that the microcanonical entropy, or Boltzmann entropy,  $s_N$ of a finite system is not necessarily real-analytic, i.\,e.\ not necessarily infinitely many times differentiable. More specifically, it was observed in \cite{CaKa06} that stationary points of the potential energy function $V_N(q)$ of a classical many-body system with continuous variables $q=(q_1,\dots,q_N)$ give rise to these nonanalyticities. Generically, with increasing $N$, the nonanalyticities appear in higher and higher derivatives of the microcanonical entropy $s_N$. Despite this ``smoothening'' of the entropy when approaching the thermodynamic limit, it was shown recently that a finite-system nonanalyticity of $s_N$ can survive the thermodynamic limit if the Hessian determinant of $V_N$, evaluated along a suitable sequence of stationary points for different system sizes $N$, goes to zero in a suitable way. This observation was used in \cite{KaSchne08,KaSchneSchrei08} to derive an analytic criterion, local in microscopic configuration space, on the basis of which the existence of phase transitions can be analyzed and, in some cases, an exact analytical expression for the phase transition energy can be derived.

In this article, the functional form of the nonanalyticities of the finite-system entropy $s_N$ is reviewed. Based on this result, we then sketch the criterion which relates the occurrence of a phase transition to the vanishing of the Hessian determinant evaluated along a sequence of stationary points. Finally, a simple strategy is discussed which permits to construct sequences of stationary points for systems of arbitrary size $N$.

\section{Nonanalyticities of the finite-system entropy}

\begin{figure}
  \psfrag{O}{$\Omega_2$}
  \psfrag{v}{$v$}
  \psfrag{v0}{\small $\!v_0$}
  \psfrag{v1}{\small $\!v_1$}
  \psfrag{v2}{\small $\!v_2$}
  \includegraphics[height=.18\textwidth,angle=270]{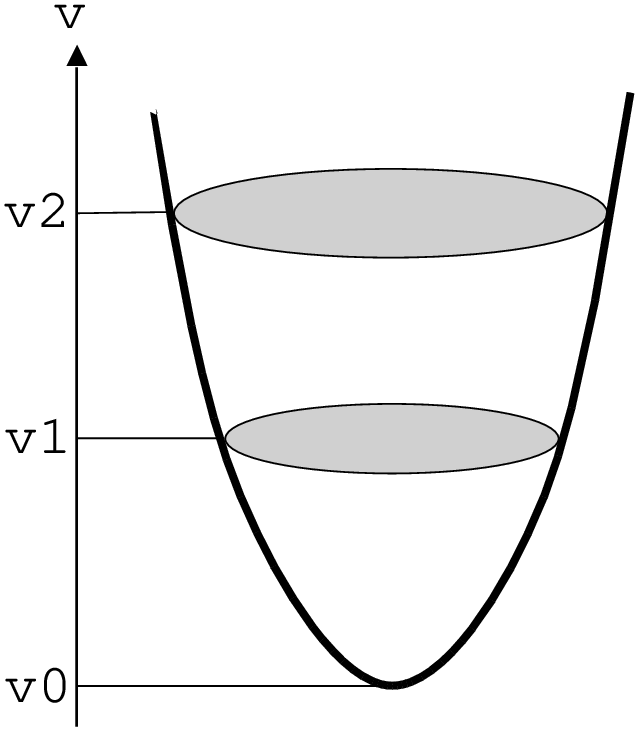}
  \hspace{0.15\textwidth}
  \includegraphics[width=.36\textwidth,angle=180]{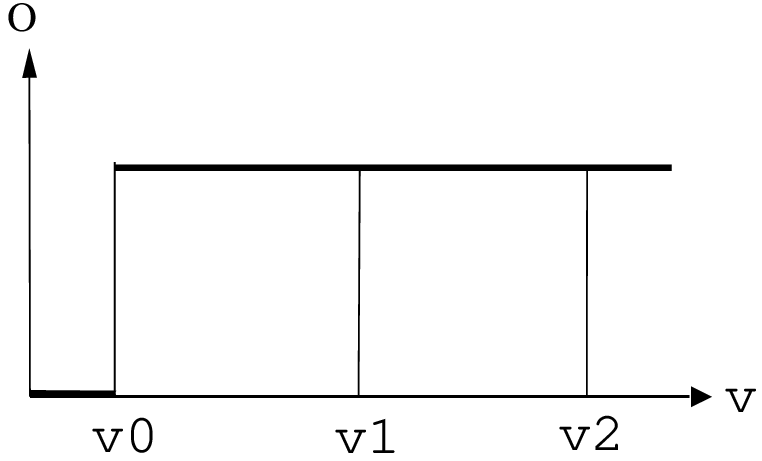}
  \caption{\label{fig:VandOmega}
As a simple example, consider a two-dimensional configurational space $\mathds{R}^2$ on which a potential energy function $V_2(q_1,q_2)=q_1^2+q_2^2$ is defined (left). The corresponding density of states as defined in \eqref{eq:Omega_N} has a discontinuity at $v=v_0$ (right).}
\end{figure}

Canonical and grandcanonical ensembles are the ones most frequently used in statistical mechanics applications. The corresponding thermodynamic potentials, i.e.\ the free energy and the grandcanonical potential, are known to be analytic functions for all finite systems sizes, and many physicists have become so used to this property that they expected all finite-system thermodynamic functions to be smooth. It is, however, fairly easy to construct counterexamples to this false expectation. The thermodynamic function we consider here is the configurational microcanonical entropy\footnote{Nonanalyticities of the ``usual'' microcanonical entropy are related to nonanalyticities of its configurational counterpart in a rather straightforward way. See \cite{CaKaNe09} for details.}
\begin{equation}
s_N(v)=\ln\Omega_N(v)/N,
\end{equation}
where
\begin{equation}\label{eq:Omega_N}
\Omega_N(v)=\int\mathrm{d} q\,\delta(V_N(q)-Nv)
\end{equation}
is the configurational density of states. The integration in \eqref{eq:Omega_N} is over configuration space, and $v$ denotes the potential energy per degree of freedom. For a potential energy function $V_2(q_1,q_2)=q_1^2+q_2^2$, the configurational density of states $\Omega_2(v)$ is easily shown to have a discontinuity at $v=v_0$ (see figure \ref{fig:VandOmega}). For a slightly less trivial example of a potential with a proper saddle point, see figure 1 of \cite{Kastner09}.

In both examples, the nonanalyticities of $\Omega_N(v)$ occur precisely at the values of the potential energy per degree of freedom $v^{\mathrm{s}}=V_N(q^{\mathrm{s}})/N$ which correspond to stationary points of the potential, i.e.\ points $q^{\mathrm{s}}$ where $\mathrm{d}V_N(q^{\mathrm{s}})=0$. This observation remains valid in higher dimensional configuration spaces, and independent of whether the stationary point is a minimum, a maximum, or a saddle point. In the following, in order to obtain a general result characterizing the nonanalytic behavior of $\Omega_N$ induced by a stationary point, we require the potential $V_N$ to be a Morse function, i.e.\ to have a non-vanishing determinant of the Hessian $\mathcal{H}_{V^{\vphantom{l}}}$ at all stationary points of $V_N$. One may argue that this is an insignificant restriction, since Morse functions form an open dense subset of the space of smooth functions \cite{Demazure} and are therefore generic. This means that, if the potential $V_N$ we are interested in is not a Morse function, we can transform it into one by adding an arbitrarily small perturbation. An important consequence of the Morse property is that all stationary points of such a function are isolated which allows us to study the effect of a {\em single}\/ stationary point on the configurational density of states \eqref{eq:Omega_N}. Such an asymptotic analysis has been reported in \cite{KaSchneSchrei07,KaSchneSchrei08}:
\begin{theorem}\label{thm:finite}
Let $V_N:G\to\mathds{R}$ be a Morse function with a single stationary point $q^{\mathrm s}$ of index $k$ in an open region $G\subset\mathds{R}^N$. Without loss of generality, we assume $V_N(q^{\mathrm s})=0$. Then there exists a polynomial $P$ of degree less than $N/2$ such that at $v=0$ the configurational density of states \eqref{eq:Omega_N} can be written in the form
\begin{equation}\label{eq:Omega_sep}
\Omega_N(v)=P(v)+\frac{h_{N,k}(v)}{\sqrt{\left|\det\left[{\mathcal H}_{V^{\vphantom{l}}}(q^{\mathrm s})\right]\right|}}+ o(v^{N/2-\epsilon})
\end{equation}
for any $\epsilon>0$. Here $\Theta$ is the Heaviside step function, $o$ denotes Landau's little-o symbol for asymptotic negligibility, and
\begin{equation}\label{eq:h_Ni}
h_{N,k}(v)=\begin{cases}
(-1)^{k/2} \,v^{(N-2)/2} \Theta(v) & \mbox{for $k$ even,}\\
(-1)^{(k+1)/2} \,v^{(N-2)/2}\,\pi^{-1}\ln|v| & \mbox{for $N$ even, $k$ odd,}\\
(-1)^{(N-k)/2} (-v)^{(N-2)/2} \Theta(-v) & \mbox{for $N,k$ odd.}
\end{cases}
\end{equation}
\end{theorem}
For a proof of this result see \cite{KaSchneSchrei08}. In short, we see from theorem \ref{thm:finite} that, at a nonanalyticity of $\Omega_N(v)$ induced by the presence of a stationary point of $V_N$, the configurational density of states is $\lfloor (N-3)/2 \rfloor$-times differentiable at the corresponding value $v^{\mathrm s}$ of the potential energy. Hence, when increasing the number $N$ of degrees of freedom, the non-differentiability occurs in higher and higher derivatives of $\Omega_N(v)$ [or $s_N(v)$], and one might naively expect the nonanalyticity to disappear in the thermodynamic limit.

\section{Flat stationary points and phase transitions}

The result of the previous section, and in particular the unbounded growth of the differentiability with increasing $N$, does not seem to suggest any connection between stationary points of the potential energy function and phase transitions in the thermodynamic limit $N\to\infty$. There is, however, a result by Franzosi and Pettini which indeed establishes such a relation. The following sloppy reformulation of the result will be sufficient for our purposes.
\begin{sloppytheorem}\label{thm:FraPe}
Let $V_N$ be the potential of a system with $N$ degrees of freedom and short-range interactions. If some interval $[a,b]$ of potential energies per degree of freedom remains, for any large enough $N$, free of stationary values of $V_N$, then the configurational entropy $s(v)=\lim_{N\to\infty}s_N(v)$ does not show a phase transition in this interval.
\end{sloppytheorem}
Note that a precise formulation of this theorem requires further technical conditions on the potential $V_N$ (see \cite{FraPe04,FraPeSpi07} for details).

Indeed, the finite-system result of theorem \ref{thm:finite} can be helpful towards an understanding of how nonanalyticities of the entropy $s_N$ can give rise to a nonanalyticity in the thermodynamic limit: From equation \eqref{eq:Omega_sep}, we observe that the nonanalytic term $h_{N,k}$ comes with a prefactor of $1/\sqrt{\left|\det\left[{\mathcal H}_{V^{\vphantom{l}}}(q^{\mathrm s})\right]\right|}$. Although, with increasing $N$, the differentiability of $h_{N,k}$ grows unboundedly, this ``smoothing'' may be counterbalanced by a vanishing (with increasing $N$) Hessian determinant. This intuition has been made rigorous in \cite{KaSchne08,KaSchneSchrei08}, but here we will give only a sloppy reformulation capturing the essence of this result.
\begin{sloppytheorem}\label{thm:flat}
The nonanalytic contributions of the stationary points of $V_N$ to the configurational entropy cannot induce a phase transition at a potential energy per particle $v_{\mathrm t}$ if, in a neighborhood of $v_{\mathrm t}$,
\begin{enumerate}
\item the number of critical points is bounded by $\exp(CN)$ with some $C>0$, and
\item the stationary points do not become ``asymptotically flat'' in the thermodynamic limit, i.e.\ $\lim_{N\to\infty}\bigl|\det \mathcal{H}_{V^{\vphantom{l}}}(q^{\mathrm{s}})\bigr|^{1/N}$ is bounded away from zero for any sequence of stationary points $q^{\mathrm{s}}$ lying in the vicinity of $v_{\mathrm t}$.
\end{enumerate}
\end{sloppytheorem}
For a precise formulation and a proof of this result see \cite{KaSchneSchrei08}. In short, this result classifies a subset of all stationary points of $V_N$ as harmless as what regards phase transitions and leaves only the asymptotically flat ones as candidates capable of causing a phase transition.

\section{Special sequences of stationary points}

Importantly for the application of sloppy theorem \ref{thm:flat}, knowledge of a suitably chosen subset of the stationary points of $V_N$ may be sufficient: If one manages to find some sequence of stationary points such that, along this sequence,
\begin{equation}
\lim_{N\to\infty}\bigl|\det \mathcal{H}_{V^{\vphantom{l}}}(q^{\mathrm{s}})\bigr|^{1/N}=0,
\end{equation}
the corresponding limiting value $v_\text{t}=\lim_{N\to\infty}V_N\bigl(q^\text{s}\bigr)/N$ is a good candidate for the exact value of the phase transition potential energy. This idea was first employed by Nardini and Casetti in \cite{NardiniCasetti09}, where suitably constructed sequences of stationary points were used to single out the phase transition of a model of gravitating masses and determine its critical energy.

To illustrate how special sequences of stationary points can be constructed, we consider a one-dimensional $XY$ model with periodic boundary conditions, characterized by the potential energy function
\begin{equation}\label{eq:Hamiltonian}
V_N(q)=\sum_{i=1}^N\sum_{j=1}^{(N-1)/2} \frac{1-\cos(q_i-q_{i+j})}{j^\alpha}
\end{equation}
where $q_i\in[-\pi,\pi)$ are angular variables, and $\alpha$ is some nonnegative exponent. For $\alpha\in[1,2]$, this model is known to show a phase transition from a ferromagnetically ordered to a paramagnetic phase, but no exact thermodynamic solution is known.

Stationary points of the potential energy \eqref{eq:Hamiltonian} have to satisfy the set of equations
\begin{equation}\label{eq:statpoints}
0=\frac{\partial V_N(q)}{\partial q_k}=\sum_{j=1}^{(N-1)/2}\frac{\sin(q_k-q_{k+j})+\sin(q_k-q_{k-j})}{j^\alpha}
\end{equation}
for $k=1,\dots,N$. To get rid of the trivial global rotational invariance of \eqref{eq:Hamiltonian}, we fix $q_N=0$ and eliminate the equation with $k=N$ in \eqref{eq:statpoints}. The thermodynamics of this reduced model is identical to that of the full one, as the contribution of one degree of freedom to the partition function is negligible in the thermodynamic limit.

\begin{figure}
  \includegraphics[width=.48\textwidth]{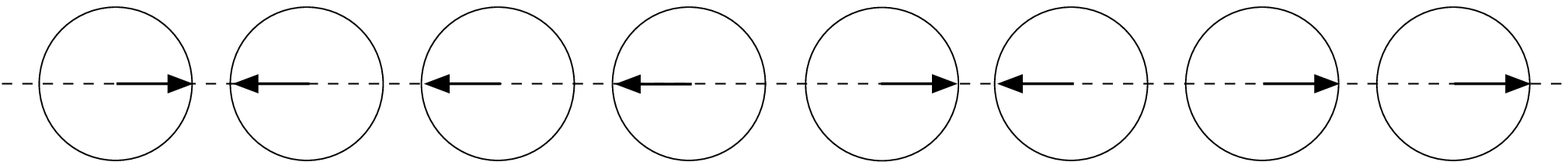}
  \hspace{.04\textwidth}
  \includegraphics[width=.48\textwidth]{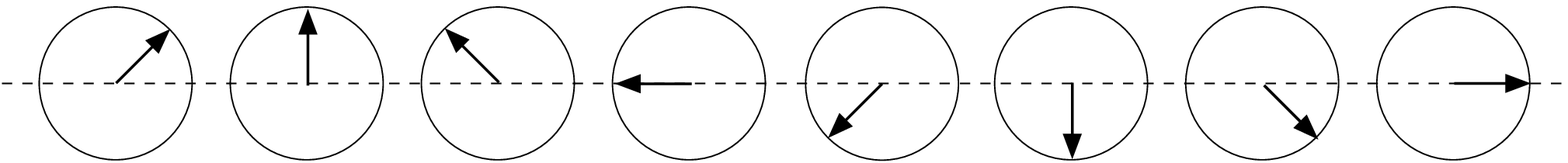}
  \caption{\label{fig:classes}
Sketch of stationary points of $V_N$ for $N=8$, where $q_i$ is the angle between the arrow and the dashed axis. Left: Stationary points where all $q_i\in\{0,\pi\}$. Right: All differences $q_k-q_{k-1}$ between neighboring angles are equal, with differences chosen such that $q_0=q_N$, in compliance with the periodic boundary conditions.
}
\end{figure}

There are two particularly simple classes of solutions of \eqref{eq:statpoints}, similar in spirit to those constructed in \cite{NardiniCasetti09} for a one-dimensional model of gravitating masses: First, any combination of $q_i\in\{0,\pi\}$ for $i=1,\dots,N-1$ will make the sine functions in \eqref{eq:statpoints} vanish. A second class of solutions is given by $q_m^{(n)}=2\pi mn/N$ for $m,n\in\{1,\dots,N\}$. These solutions have equal angles between neighboring spins. As a result, $\sin(q_k-q_{k+j})=\sin(q_{k-j}-q_k)$, and therefore each of the summands in \eqref{eq:statpoints} vanishes separately. Both classes of solutions are sketched in figure \ref{fig:classes}. To employ these classes of stationary points along the lines of sloppy theorem \ref{thm:flat}, one needs to evaluate the Hessian determinant of \eqref{eq:Hamiltonian} at the stationary points. This is work in progress and will be reported elsewhere.

\begin{theacknowledgments}
The author acknowledges financial support by the {\em Incentive Funding for Rated Researchers}\/ programme of the National Research Foundation of South Africa.
\end{theacknowledgments}

\bibliographystyle{aipproc}   

\bibliography{Granada}

\IfFileExists{\jobname.bbl}{}
 {\typeout{}
  \typeout{******************************************}
  \typeout{** Please run "bibtex \jobname" to optain}
  \typeout{** the bibliography and then re-run LaTeX}
  \typeout{** twice to fix the references!}
  \typeout{******************************************}
  \typeout{}
 }

\end{document}